\documentclass{pasj00}
\draft

\begin{document}
\SetRunningHead{T. Anada et al.}{X-ray studies of HESS J1837--069 with Suzaku and ASCA}
\Received{2008/08/29}
\Accepted{2008/10/15}

\title{X-ray studies of HESS J1837--069 with Suzaku and ASCA: a VHE
$\gamma$-ray source originated from the pulsar wind nebula}

\author{Takayasu \textsc{anada}, Ken \textsc{ebisawa}, Tadayasu {\sc
dotani}, and Aya {\sc bamba}}
\affil{Institute of Space and Astronautical Science, JAXA, 3-1-1
  Yoshinodai, Sagamihara, Kanagawa 229-8510}
\email{anada@astro.isas.jaxa.jp, ebisawa@astro.isas.jaxa.jp,
dotani@astro.isas.jaxa.jp, bamba@astro.isas.jaxa.jp}

\KeyWords{ISM: individual (HESS J1837--069) --- stars: neutron
--- X-rays: individual (AX J1838.0--0655, AX J1837.3--0652)}
\maketitle

\begin{abstract}
We present the ASCA and Suzaku studies of the TeV source HESS
 J1837--069, which has not been identified in other wave-lengths.
We confirm the presence of two X-ray sources in the Suzaku XIS image, AX
 J1838.0--0655 and AX J1837.3--0652, near both ends of the elongated TeV
 emission region. The XIS spectra of the two sources are
 reproduced by an absorbed power-law model, whose parameters are all
 consistent with those determined by the ASCA data.
 Recently, 70.5~ms X-ray pulsation has been detected with RXTE in the sky region
 including HESS J1837--069~(2008, ApJ, 681, 515).
 Using the ASCA GIS data which has both timing and imaging capabilities,
 we identified the pulsation source as AX J1838.0--0655. The pulse
 periods determined by ASCA and Suzaku,
 and that reported with RXTE indicate
 steady spin-down at $\dot{P} = 4.917(4) \times 10^{-14}$~s~s$^{-1}$.
These results suggest that AX J1838.0--0655 is an intrinsically stable
 source, and presumably a pulsar wind nebula (PWN).
We discuss the possibility that AX J1838.0--0655 is associated with HESS
 J1837--069 and the VHE $\gamma$-ray emission is originated from the PWN.
\end{abstract}

\section{Introduction}

Galactic plane survey with the H.E.S.S. Cherenkov telescope system
revealed the presence of dozens of new very-high-energy (VHE) $\gamma$-ray
sources~\citep{2005Sci...307.1938A,2006ApJ...636..777A}. Many of them
have no counterparts in other wave-lengths, and are thus called
``unidentified (unID) TeV sources''.
Today, there are about 40 such unidentified TeV sources on the
Galactic plane~\citep{2007arXiv0712.3352H}. Most of them are located
within 1 degree from the Galactic plane, and some are intrinsically extended.
Some of them are suspected to be pulsar wind nebulae (PWNe) or supernova
remnants (SNRs)\citep{2007arXiv0712.3352H}, but origins of most sources
are still unclear.

HESS J1837--069 is one of the VHE $\gamma$-ray sources discovered by the
Galactic survey with H.E.S.S. in
2004 centering at the position of (RA, Dec) = (\timeform{18h37m42s.7},
\timeform{-6D55'39''}). It has a significantly elongated shape with an
extension of
\timeform{7'}$\times$\timeform{3'}~\citep{2005Sci...307.1938A,2006ApJ...636..777A}.
No X-ray sources are known to be positionally coincident to the center
of HESS
J1837--069, although AX J1838.0--0655, which is located at the Galactic
south edge of the HESS source, was suggested to be a possible
counterpart~\citep{2008ApJ...681..515G}.
ASCA observation revealed that AX J1838.0--0655 has very hard and strongly
absorbed
spectrum~\citep{2003ApJ...589..253B}.
INTEGRAL~\citep{2005ApJ...630L.157M} observations also support this result. 
Recently, \citet{2008ApJ...681..515G} discovered a 70.5~ms pulsation with
RXTE in the sky field including AX J1838.0--0655, and also resolved a
bright point source surrounded by diffuse emission with Chandra.
They concluded that AX J1838.0--0655 is a PWN.
Here we report the results of ASCA archival data analysis and the newly
obtained Suzaku observation of HESS J1837--069/AX J1838.0--0655.

\section{Observation}
\label{sec:observation}

We observed HESS J1837--069 with
Suzaku~\citep{2007PASJ...59S...1M} in March, 2007.
Suzaku is equipped with two types of the instruments: the X-ray Imaging Spectrometer
(XIS:~\cite{2007PASJ...59S..23K}) at the focal plane of X-Ray Telescope
(XRT:~\cite{2007PASJ...59S...9S}) and the Hard X-ray Detector
(HXD:~\cite{2007PASJ...59S..35T}, \cite{2007PASJ...59S..53K}), which is
non-imaging instrument with a \timeform{34'}$\times$\timeform{34'}
full-width half-maximum (FWHM) square field-of-view~(FOV) below $\sim$ 100~keV.
The observation was carried out with the HXD optical axis (which is
$\sim$ \timeform{3'} offset of that of XIS) placed at
the center of HESS J1837--069 in order to optimize the HXD throughput.
Three XISs (XIS 0, 1, 3) were operated in the normal clocking mode with
Spaced-row Charge Injection (SCI)~\citep{2008PASJ...60S...1N}. 
We analyzed the data prepared by the version 2.0 pipeline.
We applied the standard screening criteria to both the
XIS\footnote{http://www.astro.isas.jaxa.jp/suzaku/process/v2changes/criteria\_xis.html}
and
HXD\footnote{http://www.astro.isas.jaxa.jp/suzaku/process/v2changes/criteria\_hxd.html}
data to obtain cleaned event lists.
After the data screening, the net exposures were 42.2~ks and 37.7~ks for XIS
and HXD, respectively. 

We also used the same ASCA archival data in the current study, that were
previously published by \citet{2003ApJ...589..253B}.
HESS J1837--069 was in the FOV of ASCA GIS in
both the 1997 and 1999 observations~\citep{2003ApJ...589..253B}.
The data was screened based on the same criteria as
\citet{2003ApJ...589..253B}.

The net exposures after screening are summarized in
table~\ref{tbl:obslog} with the journal of Suzaku and ASCA observations.
We used HEADAS version 6.3.1 software package for all the data analysis
in the present paper.

\begin{table*}
 \begin{center}
 \caption{Journal of the ASCA/Suzaku observations of HESS J1837--069}\label{tbl:obslog}
  \begin{tabular}{ccccccc}
   \hline
   Satellite & Sequence ID & Start time (UT) & End time (UT) &
   \multicolumn{2}{c}{Aim point (J2000)} &
   Net exposure \\
   & & \multicolumn{2}{c}{(yyyy/mm/dd hh:mm)} & R.A. &
   Decl. & (ks) \\
   \hline
   ASCA & 55002090 & 1997/10/14 09:12 & 1997/10/14 14:17 &
   \timeform{18h37m48s.0} & \timeform{-6D36'42''} &
   12.4\footnotemark[$*$] / 8.4\footnotemark[$\dagger$] \\
   ASCA & 57029000 & 1999/09/26 18:12 & 1999/09/28 03:30 & \timeform{18h37m45s.6}
   & \timeform{-6D36'45''} &
   37.3\footnotemark[$*$] / 17.6\footnotemark[$\dagger$] \\
   Suzaku & 401026010 & 2007/03/05 12:49 & 2007/03/06 10:17 & \timeform{18h37m42s.7}
   & \timeform{-6D55'39''} & 42.2\footnotemark[$\ddagger$] / 37.7\footnotemark[$\S$] \\
   \hline
   \multicolumn{7}{@{}l@{}}{\hbox to 0pt{\parbox{180mm}{\footnotesize
   \footnotemark[$*$] GIS, high and medium bit-rate data
   \par\noindent
   \footnotemark[$\dagger$] GIS, high bit-rate data only
   \par\noindent
   \footnotemark[$\ddagger$] XIS
   \par\noindent
   \footnotemark[$\S$] HXD-PIN
   }\hss}}
  \end{tabular}
 \end{center}
\end{table*}

\section{Results}
\label{sec:results}

\subsection{X-ray Image}
\label{sec:image}

\begin{figure}
 \begin{center}
  \FigureFile(80mm,80mm){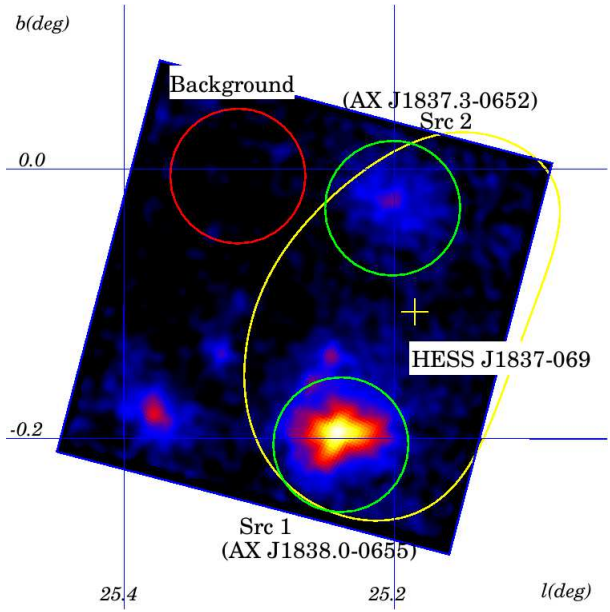} 
 \end{center}
 \caption{
 Suzaku XIS (0+1+3) image in Galactic coordinate around HESS J1837--069 in the
 0.4-10.0 keV band. The data between 5.73--6.67~keV were filtered out to
 remove the calibration sources. The pseudo-color represents
 vignetting-corrected, log-scaled intensity levels.
 Yellow line represents the $e^{-1/2}$ contour to the peak
 of the VHE $\gamma$-ray image of HESS J1837--069\footnotemark~\citep{2006ApJ...636..777A}.
 Two sources in the green circles are referred to Src 1 (AX
 J1838.0--0655) and Src 2 (AX J1837.3--0652). Background data were
 extracted from the red circle whose vignetting is almost the same as
 the two sources. All the radii of the circles are 
 \timeform{3'}. 2E1835.5-0650 was also detected at the position of $(l,
 b) = (\timeform{25D.38}, \timeform{-0D.18})$.
}
 \label{fig:img_reg-hess}
\end{figure}

We extracted the XIS image in 0.4--10.0~keV for each sensor. The data
between 5.73--6.67~keV were removed from the image to
exclude the calibration sources.
We corrected the vignetting effect by dividing the image with the flat
sky image simulated at 4.0~keV using the XRT+XIS simulator {\tt
xissim}~\citep{2007PASJ...59S.113I}. 
The image was binned to 8$\times$8 pixels and smoothed with a
Gaussian function of $\sigma =\timeform{0'.42}$.
Combined Suzaku XIS (0+1+3) image is shown in
figure~\ref{fig:img_reg-hess}.
Two X-ray sources are seen at the edges of galactic south
and north of HESS J1837--069. Hereafter, we refer the two sources as Src
1 and Src 2 in this paper as indicated in figure~\ref{fig:img_reg-hess}.
We determined the peak positions of Src 1 and Src 2 as listed in
table~\ref{tbl:position}.
These positions are consistent with those of the ASCA 
sources, AX J1838.0--0655 and AX J1837.3--0652, respectively.
Locations of these two sources are significantly offset from the center
of HESS J1837--069 (\timeform{6'.4} for Src 1 and \timeform{5'.7} for
Src 2), although they are both spatially compatible with the reported
extension of HESS J1837--069.

\begin{table}
 \begin{center}
 \caption{Suzaku detected X-ray sources in the vicinity of HESS J1837--069}\label{tbl:position}
  \begin{tabular}{cccc}
   \hline
   Src & \multicolumn{2}{c}{coordinate\footnotemark[$*$] (J2000)} & Association \\
   & R.A. & Decl. &\\
   \hline
   1 & \timeform{18h38m03s} & \timeform{$-$06D55'43''} & AX J1838.0--0655 \\
   2 & \timeform{18h37m21s} & \timeform{$-$06D53'14''} & AX J1837.3--0652 \\
   \hline
   \multicolumn{4}{@{}l@{}}{\hbox to 0pt{\parbox{85mm}{\footnotesize
   \footnotemark[$*$] Error radius (90\%) is \timeform{19''}~\citep{2008PASJ...60S..35U}
   }\hss}}
  \end{tabular}
 \end{center}
\end{table}

\footnotetext{Fits file is available from
 http://www.mpi-hd.mpg.de/hfm/HESS/public/publications/ApJ\_636.html}

\subsection{Energy Spectra}
\label{sec:spec}

\begin{figure}
 \begin{center}
  \FigureFile(80mm,80mm){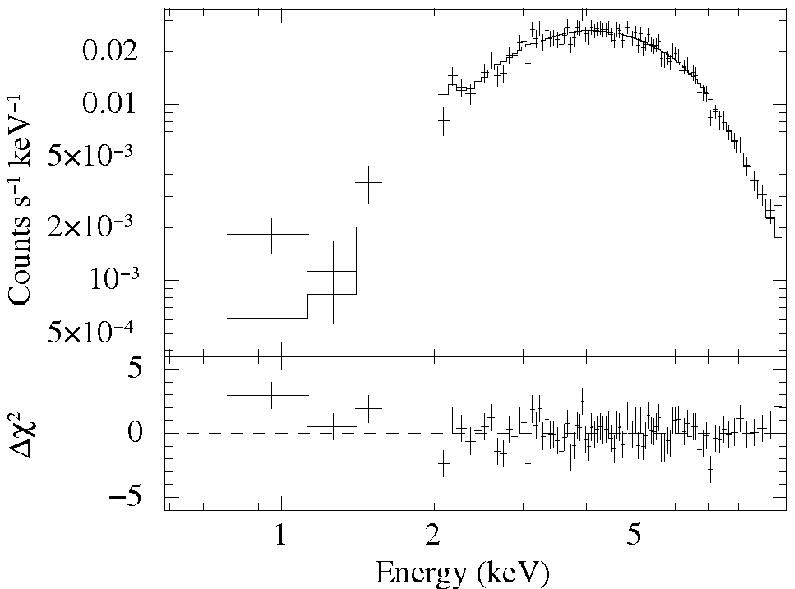} 
  \hspace{0.5cm}
  \FigureFile(80mm,80mm){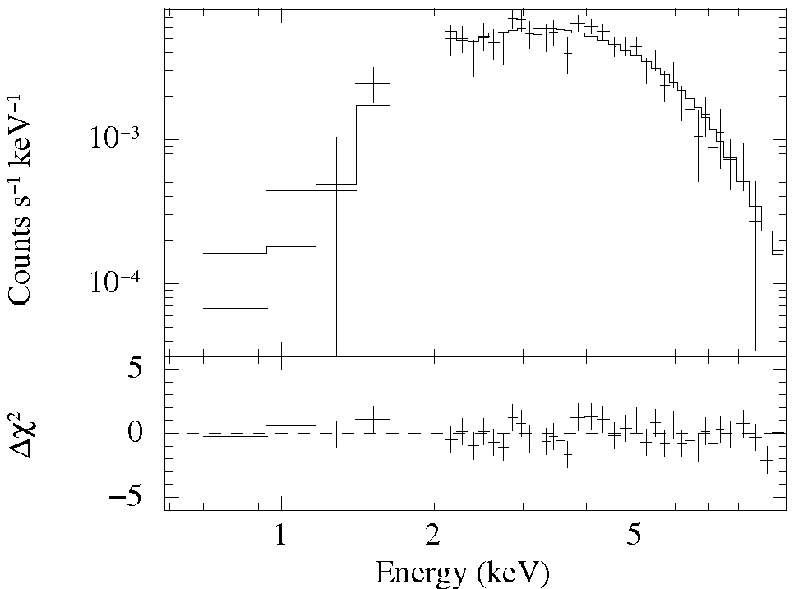} 
 \end{center}
 \caption{Background-subtracted XIS spectra of Src 1 (left)
 and Src 2 (right), respectively. Solid lines show the
 best-fit model. The bottom panels show residuals to the best-fit model.}
 \label{fig:spectrum_sminusb}
\end{figure}

We extracted the energy spectra of Src 1 and Src 2
within the \timeform{3'} circular regions centered on the
sources to avoid nearby faint point source
detected by Chandra, which enclose $\sim$90\% of photons for each point source (green
circles in figure~\ref{fig:img_reg-hess}). Background
data were extracted from the source-free region as indicated by a red
circle in figure~\ref{fig:img_reg-hess}, whose vignetting is almost same
as that of Src 1 and Src 2.
Figure~\ref{fig:spectrum_sminusb} shows the XIS spectra (averaged for XIS 0,
1 and 3) of Src 1 and Src 2 in the 0.4--10~keV band. The 1.7--2.0~keV band was
ignored from the analysis because of large calibration uncertainties around
the Si edge\footnote{http://www.astro.isas.jaxa.jp/suzaku/doc/suzaku\_td/}.
We generated detector and auxiliary response files using {\tt
xisrmfgen} and {\tt xissimarfgen} for each source~\citep{2007PASJ...59S.113I}.
We fitted the spectra with an
absorbed power-law model using XSPEC version 12.3.1. The fit results are
summarized in table~\ref{tbl:fit}.

We also reanalyzed the data obtained with ASCA GIS in 1997 and
1999 to determined the errors of the source fluxes, which are not published
in \citet{2003ApJ...589..253B}. We extracted the data of Src 1 and
Src 2 in the 0.7--10~keV band from the same regions
as~\citet{2003ApJ...589..253B} and subtracted the background extracted
from the source-free regions near the sources.
We summed up the data in 1997 and 1999 to improve statistics as was done
by \citet{2003ApJ...589..253B}.
We fitted the spectra with an absorbed power-law model and
calculated the single-parameter 90\% confidence regions of the
unabsorbed fluxes. The results are listed in table~\ref{tbl:fit}.

\begin{table*}
 \begin{center}
 \caption{Best-fit parameters of the absorbed power-law model}
 \label{tbl:fit}
  \begin{tabular}{cccccc}
   \hline
   Source & Instrument & $\Gamma$ & $N_{\mathrm{H}}$\footnotemark[$*$] &
   Flux\footnotemark[$\dagger$] & $\chi^2$ / dof \\
   \hline
   1  & Suzaku/XIS & 1.27$\pm0.11$  & 5.4$\pm0.5$ & 13.2$^{+0.8}_{-0.6}$ & 97.7 / 86 \\
     & ASCA/GIS & 0.8$\pm0.4$ & 4.0$^{+1.6}_{-1.3}$
   & 14$^{+2}_{-1}$ & 12.9 / 12 \\
   2  & Suzaku/XIS & 2.2$^{+0.4}_{-0.3}$ & 4.6$^{+1.5}_{-1.0}$ & 3.1$^{+1.5}_{-0.7}$ & 27.1 / 37 \\
    & ASCA/GIS & 2.4$^{+0.9}_{-0.8}$ & 6.2$^{+3.4}_{-2.2}$
   & 6$^{+13}_{-3}$ & 35.6 / 33 \\
   \hline
   \multicolumn{6}{@{}l@{}}{\hbox to 0pt{\parbox{180mm}{\footnotesize
   Note. --- Errors represent single-parameter 90\% confidence limit.
   \par\noindent
   \footnotemark[$*$] Absorption column density using the solar
   abundance ratio~\citep{1982GeCoA..46.2363A} in the unit of $10^{22}$~cm$^{-2}$.
   \par\noindent
   \footnotemark[$\dagger$] Unabsorbed flux in the 0.7-10.0~keV band in the
   unit of 10$^{-12}$~ergs~cm$^{-2}$~s$^{-1}$.
   }\hss}}
  \end{tabular}
 \end{center}
\end{table*}

\subsection{Timing Analysis}
\label{sec:timing}

Because X-ray pulsation was recently detected from the sky region
including HESS J1837--069~\citep{2008ApJ...681..515G}, we carried out
timing analysis of the Suzaku HXD-PIN and ASCA GIS data to study the
long-term change of the pulse period and to identify the source of pulsation.
Although HXD-PIN has a large FOV, there are no significant contamination 
sources to Src 1 in its FOV. Thus we searched for pulsation with HXD-PIN.
When we analyzed the HXD-PIN data, we carefully examined the best energy
range to maximize the signal-to-noise ratio.
In order to minimize the contribution of non-X-ray background
(NXB)\footnote{We used the quick-version of the NXB model
explained in SUZAKUMEMO-2008-03 \\
(http://www.astro.isas.jaxa.jp/suzaku/doc/suzakumemo/suzakumemo-2008-03.pdf),
and also in \citet{2009PASJ}.}
, cosmic X-ray background (CXB) and Galactic ridge X-ray emission
(GRXE), we selected 12--23~keV for the analysis of the PIN data.
In this energy band, the source count rate was 0.41~counts~s$^{-1}$ whereas the
count rates of NXB and CXB+GRXE were 0.32 and 0.04~counts~s$^{-1}$, respectively.
We then applied barycentric correction to the PIN data using {\tt
aebarycen}~\citep{2008PASJ...60S..25T}.
Using the {\tt efsearch} ftool, we searched for pulsation at 128 trial
periods between 0.0704949--0.0704987~sec and found a significant
peak at 0.07049672(8)~s with $\chi^2 \sim 46$ (9 degrees of freedom).
Here the 1$\sigma$ error is indicated in parentheses, which was
calculated according to~\citet{1996A&AS..117..197L}. The chance
probability to obtain such a large $\chi^2$ in 128 trial is only 0.01\%.

Whereas HXD-PIN is a non-imaging instrument, GIS has an imaging
capability and can extract the events from the source region.
Assuming that the pulsation originates from Src 1, we
searched the GIS data of Src 1 for pulsation. Although the
nominal time resolution of GIS in the PH mode is 62.5~ms in the
high telemetry bit rate~\citep{1996PASJ...48..157O}, we can achieve
higher time resolution up to 4~ms when the total count rate is
low.
The GIS event data were output to the fixed format telemetry, whose
relative location within a 62.5~ms slot was designed to indicate the
finer timing
information\footnote{http://heasarc.gsfc.nasa.gov/docs/asca/newsletters/time\_assignment4.html}.
The photon arrival time of GIS in the ASCA
archive is all assigned taking this finer timing information into
account. Because Src 1 (and other sources in the GIS FOV) is dim
(2.3~counts~s$^{-1}$ compared to the telemetry capacity of 128
counts~s$^{-1}$), we can fully
utilize the higher than the nominal time resolution of GIS.
The GIS image of Src 1 was elongated due to the proximity to
the edge of the FOV. Thus we used an elliptical region of
\timeform{1'.7}$\times$\timeform{0'.9} to extract the source events. We
used only the high bit rate data in 1999, and 224 photons in 2--10~keV
(including background) was extracted in total. We applied barycentric correction
using {\tt timeconv}.
We searched for pulsation at 128 trial periods between
0.0704838--0.0704865~sec and found a significant peak at 0.070485(2)~sec
with $\chi^2 \sim 38$ (9 degrees of freedom). The chance probability to
obtain a $\chi^2$ larger than 38 is only 0.2\%. Thus the peak is
statistically significant.

\begin{figure}
 \begin{center}
  \FigureFile(80mm,80mm){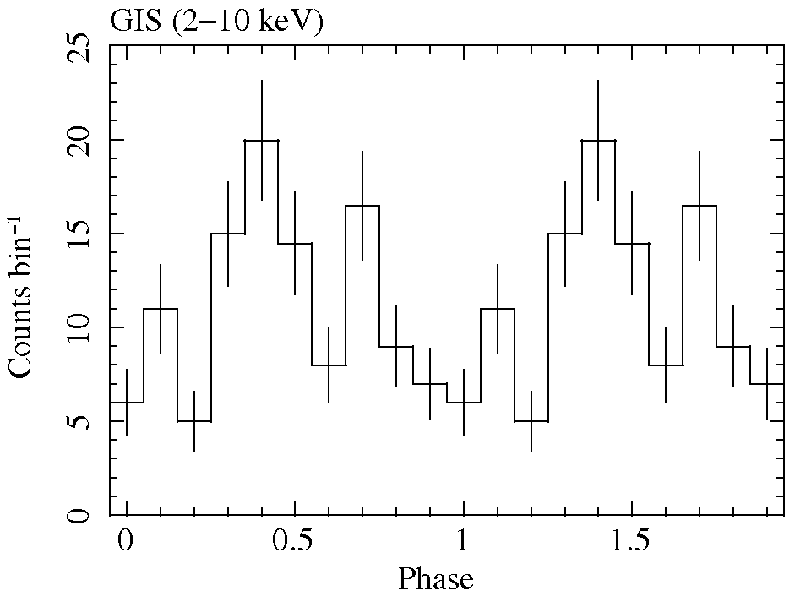}
  \hspace{0.5cm}
  \FigureFile(80mm,80mm){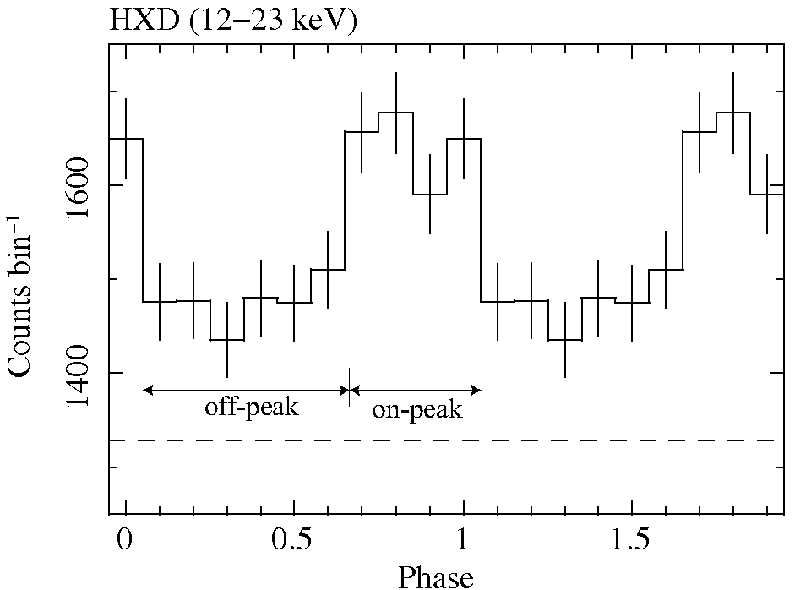}
 \end{center}
 \caption{Left: Pulse profile obtained with the GIS data of Src 1 in the
 2--10~keV band. Right: Pulse profile obtained with PIN in the
 12--23~keV band. Dashed line indicates the estimated count rate of
 the background. The pulse profile is corrected for the constant dead
 time (93\%).}
 \label{fig:efold}
\end{figure}

\begin{figure}
 \begin{center}
  \FigureFile(80mm,80mm){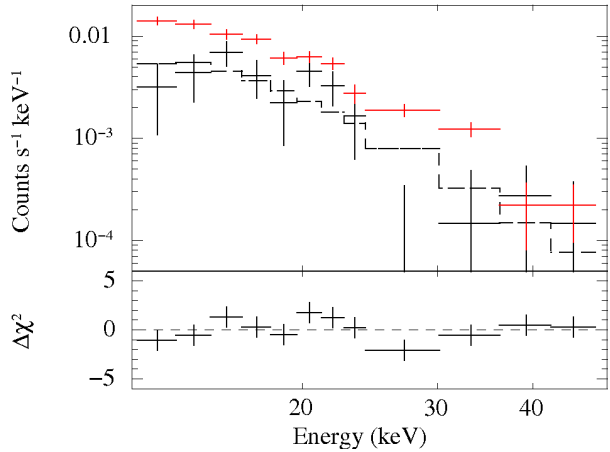}
 \end{center}
 \caption{Black line represents HXD-PIN pulsed spectrum obtained by
 subtracting the off-peak spectrum from the on-peak spectrum. Dashed
 histogram represents the best-fit power-law model. Total HXD-PIN
 spectrum subtracted NXB is also shown as red line.}
 \label{fig:on-off}
\end{figure}

We show the folded pulse profiles of PIN and GIS in
figure~\ref{fig:efold} with respective pulse periods determined above.
Parameters determined by the timing analysis above are summarized in
table~\ref{tbl:timingpara}.
We divided the PIN data into two phases: on-peak (phase 0.65--1.05)
and off-peak (phase 0.05-0.65).
We extracted the pulsed spectrum by subtracting the off-peak spectrum from the
on-peak spectrum, and fitted it with a power-law model in the
12--50~keV band. The fitted 
spectrum is shown in figure~\ref{fig:on-off}. Best-fit photon index and
the flux in the 12--50~keV band were found to be $\Gamma =
2.0^{+1.0}_{-0.9}$ and $F$ = $1.8^{+0.8}_{-0.6}\times
10^{-11}$~ergs~cm$^{-2}$~s$^{-1}$. 
We tried the same analysis for the GIS data, but could not obtain a
meaningful result due to poor statistics.

\begin{table}
 \begin{center}
 \caption{Timing Parameters of Src 1}\label{tbl:timingpara}
  \begin{tabular}{lcc}
   \hline
   Parameter & GIS & HXD-PIN \\
   \hline
   Epoch (MJD TDB) & 51447 & 54164 \\
   Period, $P$ (ms) & 70.485(2) & 70.49672(8) \\
   \hline
   \multicolumn{3}{@{}l@{}}{\hbox to 0pt{\parbox{85mm}{\footnotesize
   Note. --- 1$\sigma$ uncertainties are given in parentheses.
   }\hss}}
  \end{tabular}
 \end{center}
\end{table}

\section{Discussion}
\label{sec:discussion}

\subsection{Nature of Src 1}
\label{sec:axj1838}

\begin{figure}
 \begin{center}
  \FigureFile(80mm,80mm){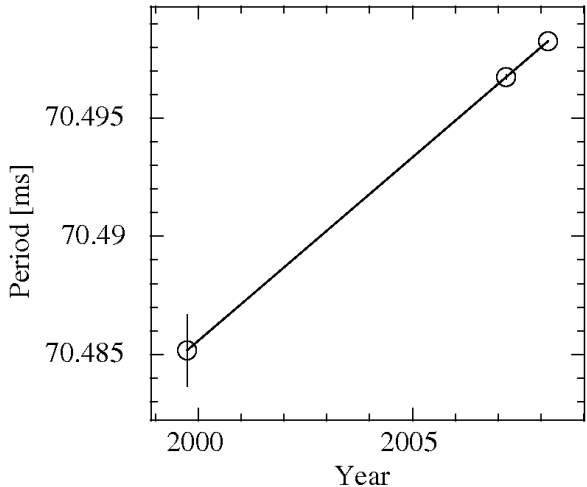} 
 \end{center}
 \caption{History of the pulse period. The data points from left to
 right indicate the pulse periods determined by ASCA, Suzaku and RXTE, respectively.}
 \label{fig:time_period}
\end{figure}

We have confirmed that the source of 70.5~ms pulsation, which was
recently discovered by RXTE, is indeed Src 1. The pulse profiles
obtained by GIS and HXD-PIN are consistent with that of RXTE.
We show in figure~\ref{fig:time_period} a long-term history of the pulse
period.
All the data are consistent with the stable spin-down with
$\dot{P} = 4.917(4) \times 10^{-14}$~s~s$^{-1}$ (1$\sigma$ error is
indicated in parentheses), which is also consistent with the instantaneous
spin-down rate determined by RXTE for the interval of $\sim$ 16
days~\citep{2008ApJ...681..515G}.
The spectral parameters (flux, photon index, column density) of Src 1
determined by the GIS data were all consistent with those of XIS. 
The flux and the column density were also consistent with those of Chandra.
Therefore, we conclude that Src 1 is an intrinsically stable source.

\citet{2008ApJ...681..515G} claimed that the Chandra spectrum of the
pulsar component was significantly harder ($\Gamma = 0.5\pm0.2$) than
the pulsed component obtained by RXTE in 2--20~keV
($\Gamma=1.2\pm0.1$). They suggested a steepening of the spectrum in the
8--15~keV band.
The Suzaku HXD-PIN data indicate a photon index of
$\Gamma=2.0^{+1.0}_{-0.9}$ for the pulsed component in the 12--50~keV.
This photon index is consistent with the RXTE result and supports the
steepening of the spectrum in the hard X-ray band.

\subsection{Nature of Src 2}
\label{sec:axj1837}

Src 2 was also observed by ASCA and is known to have a non-thermal
spectrum~\citep{2003ApJ...589..253B}. Current Suzaku observation 
suggests no time variation of Src 2 since the ASCA
observation. Therefore we speculate that Src 2
is also a PWN. However, Chandra could detect neither the
putative pulsar point source in Src 2 nor any
pulsation~\citep{2008ApJ...681..515G}. Further observations are
required to reveal the nature of this source.

\subsection{Origin of HESS J1837--069}
\label{hessj1837}

Because Src 1 is slightly offset from the center of
HESS J1837--069, we need to check the chance coincidence of the sources.
The chance probability for high-power pulsars and VHE $\gamma$-ray
sources was estimated by \citet{2007arXiv0709.4094C}. For Src 1 with
$\dot{E}/d^2 \sim 10^{35}$~ergs~s$^{-1}$~kpc$^{-2}$, the chance
probability is $\sim12 \%$. Thus we consider Src 1 is likely to be associated to
HESS J1837--069.

\bigskip
We thank Gerd P{\"u}hlhofer and Stefan Wagner for providing the HESS
image FITS file of HESS J1837--069. We also thank Ryo Yamazaki for his
valuable comment on the discussion.
This work made use of archival data of ASCA from Data Archives and
Transmission System (DARTS) maintained by C-SODA at JAXA/ISAS.
This work was partially supported in part by
Grant-in-Aid for Scientific Research
of the Japanese Ministry of Education, Culture, Sports, Science
and Technology, No.~19$\cdot$4014 (A.~B.).

\end{document}